# Platform for Assessing Strategic Alignment Using Enterprise Architecture: Application to E-Government Process Assessment

Kaoutar Elhari and Bouchaib Bounabat

**Al-Qualsadi Research & Development Team, National Higher School for Computer Science and System analysis (ENSIAS),
Mohammed V$_{th}$ University-Souissi
Rabat, Mohammed Ben Abdallah Regragui avenue, Madinat Al Irfane, BP 713, Agdal, Morocco**
elhari_kaoutar@yahoo.fr , bounabat@ensias.ma

**Abstract**
This paper presents an overview of S2AEA (v2) (Strategic Alignment Assessment based on Enterprise Architecture (version2)), a platform for modelling enterprise architecture and for assessing strategic alignment based on internal enterprise architecture metrics. The idea of the platform is based on the fact that enterprise architecture provides a structure for business processes and information systems that supports them. This structure can be used to measure the degree of consistency between business strategies and information systems. In that sense, this paper presents a platform illustrating the role of enterprise architecture in the strategic alignment assessment. This assessment can be used in auditing information systems. The platform is applied to assess an e-government process.

## 1. Introduction

The information technology investment impacts positively on business performance. In order to reach a good impact, IT must constantly be appropriated to the business strategy. The strategic alignment (SA) has been studied since 1993 [1] how to coordinate the company's strategy with the information system strategy in order to improve the efficiency of information systems which support the company's business. Indeed, misaligned solutions have negative effects on the business level and, in turn, can reduce the value of services provided by the company.

On the other hand, the concept of enterprise architecture has come, more than twenty years ago, to address two problems: systems complexity and poor strategic alignment [2]. The enterprise architecture is the best way of representing information as a model illustrating the links between strategy, business and information systems [3].

Thus, this article presents a platform which assesses SA using the enterprise architecture. It is based on a set of metrics collected from several researches, classified according to the links between the layered structures proposed by enterprise architecture. The platform helps architects to improve the SA maturity level by (a) analyzing the structure of enterprise architecture and (b) suggesting the effort to do in order to reach a better level.

This article uses many concepts of [4]. It is recommended to read it before.

The layout of this paper is as follows. The second section is devoted to EA and SA concepts; the third section presents an e-government process which will be used as an example to illustrate the platform functionalities. Finally, the fourth section presents the platform developed to support SA assessment by comparing the two versions of the platform. The conclusion and future work are presented in Section 5.

## 2. Strategic Alignment Evaluation

Many terms are used in the literature to refer to the SA [5]. Thus, a lot of synonymous of alignment are proposed: congruence, harmony, correspondence, coherence, and so on. The diversity of terms used involves the diversity of meaning given to the SA concept. [5] defines it as the correspondence between a set of components (e.g. between business process and system that supports them). [6] sees it as the act of applying information technology in harmony with the strategies, needs and objectives of the business. Some others study it as the harmony between architecture and software architecture of business processes [7]. Others consider the alignment between information systems and its environment [8]. And yet others are interested in aligning business processes and systems supporting these processes [9], [10].

In this article, we study the SA as harmony or correspondence between the company strategy represented by business processes and the systems supporting them.

### 2.1 Strategic alignment evaluation

Luftman proposes a framework for measuring the alignment between a company's strategies and the information technology strategies [11]. This framework is based on the foundations of CMM (Capability Maturity Model). He proposed five levels of maturity from 1 (not alignment) to 5 (strong line). To evaluate SA in [11], six criteria were studied: communication, competency, governance, partnership, scope, architecture and skills.

[9] suggests an alignment strategy corresponding to a sequence of activities (represented by UML activity diagram). One of these activities is the evaluation of alignment. He proposed two metrics for this evaluation: Technological Coverage and Technological Adequacy. These two metrics are insufficient to assess the SA.

[12] proposes a framework for measuring alignment using a set of metrics classifying them according to four categories: intentional alignment, information alignment, functional alignment and dynamic alignment.

The purpose of this article is to assess the strategic alignment based on enterprise architecture

## 2.2 Strategic alignment evaluation based on enterprise architecture concepts

Enterprise architecture describes the enterprise structure. It represents all aggregate artifacts that are relevant to a company. There are many frameworks used to describe enterprise architecture such as [13], [14], [15] etc. But, it is often modelled as a layered organisation. The layers that are usually recognised in this context are the business layer, the application layer, the information layer and the technology layer.

The definitions given to different layers in this paper are:

• The business layer represents the business of the company which is represented by a set of processes. Each process may consist of several activities (or sub processes). The processes or activities are supported by applications and use information entities. A process is characterized by its criticity.

• The application layer represents the application layer that automates the processes and activities. Each application has functionalities that meet the needs of the business processes. An application is described by a set of quality factors defined by [16]:

• The information layer is the data layer which is represented by information entities that can be found in data sources and which are formed by attributes. An attribute can be described by several qualifiers: secure, confidential, redundant.

• The technology layer is the layer of technical infrastructure including operating systems and technologies.

Figure 1 presents the metamodel used in this paper using a UML class diagram.

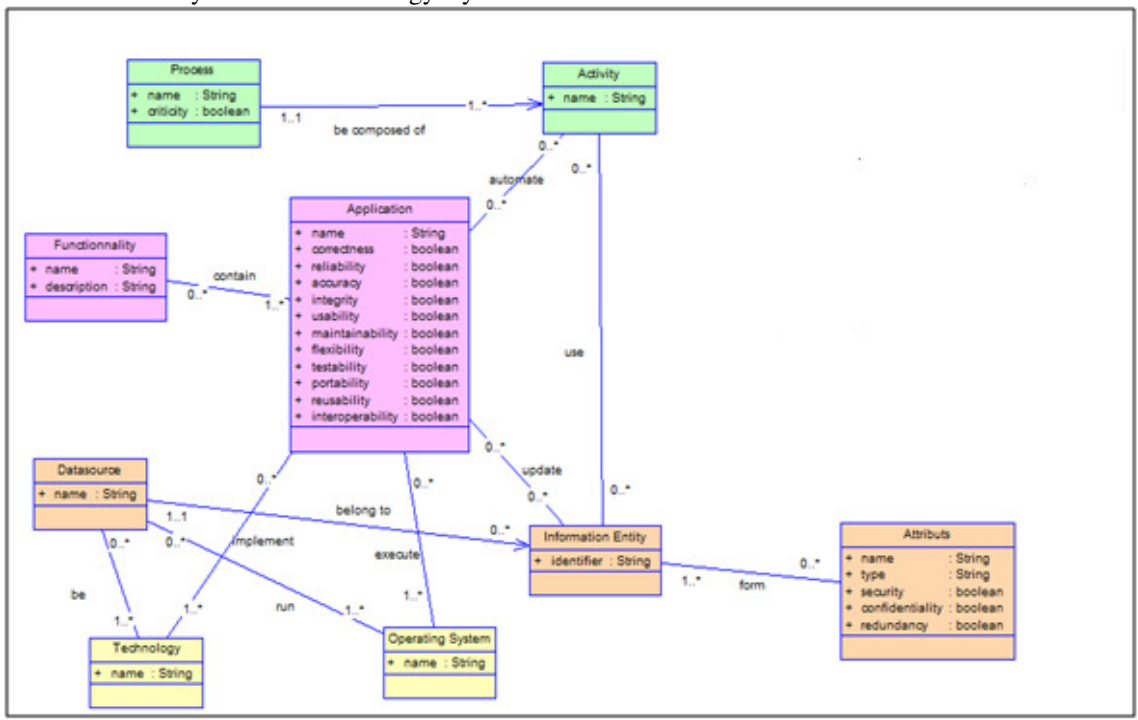

Fig1: Enterprise architecture metamodel

Many authors such as [17], [18], [19], [20], [21], [22], have associated the EA and SA concepts.

In this article, we are interested in detailed assessment of SA by examining the links between the various enterprise architecture layers.

Thus [19] develops an assessment of the SA from the links between the different EA layers, especially the business-application link and the business-data link. The metrics used in [19] use the quality criteria proposed by [16] on software quality and the notion of critical business processes that means a business process priority, which contributes to specific goals within the company and which is not superfluous [19].

Furthermore, studies such as [20], [21] present metrics for assessing the information system architecture.

In the same sense, [22] proposes a model of business of non-alignment with the information system by comparing it to medical science approaches. Thus, the authors suggest a set of cases where the business is not aligned with the information system. Then they present for everyone the organ system of the non alignment, symptoms, signs, syndromes and their etiologies. Then they suggest a diagnosis, therapy and prophylaxis.

The authors in [4] propose a strategic alignment maturity model based on enterprise architecture. The authors collected a set of metrics from several researches for each enterprise architecture internal link. They use the enterprise architecture metamodel presented in Figure1 and develop an evaluation tool for strategic alignment maturity which calculates metrics values and infers the maturity level for each layer's link. They propose five levels (chaotic, poor, average, good, very good). Their approach is represented on the diagram of the figure 2.

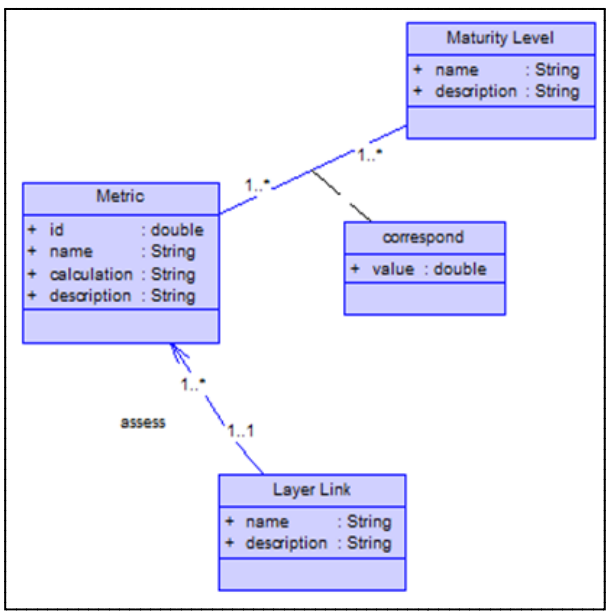

Fig2: maturity model diagram

This paper presents a platform supporting enterprise architecture modelling, calculating metrics values and proposing where architectures have to change in order to enhance strategic alignment level. The platform is applied to an e-government process.

## 3. An E-Government case study

The process that will illustrate the assessment of strategic alignment in this article is surveys data production by using the characters automatic recognition. It was used in Morocco for the first time in the Census of Population and Housing 2004 [23]. Now, several surveys use the same process.

The process of data production using automatic characters recognition consists of several activities:

We are going to illustrate the assessment of strategic alignment by the process: data capture which is used to produce data from questionnaires of the surveys.

Data Capture process contains 7 activities:

Activity 1: Receiving of questionnaires - The first step is to receive batches of questionnaires with an electronic file that indicates the identification number of each batch.

Activity 2: Scanning - It consists of scanning documents. Its aim is to computerize paper documents to enable and prepare the automatic optical recognition.

Activity 3: Character recognizing - It translates a group of points of a scanned image into characters readable by computer programs. It uses OCR (optical character recognition) technology.

Activity 4: Key correction and coding -The objective of this activity is to monitor, validate or correct the fields that were not recognized by the OCR with a sufficient confidence level or which have a coherence formula that indicates a suspicion of error.

Activity 5: Inter-questionnaires control and correction - This process was undertaken for each batch to verify that all questionnaires within a statistical area had been processed.

Activity 6: Quality control - The objective is to verify if the number of fields misread or misinterpreted in a document's batch is not above the targets set for production. The quality control method used was implemented to produce data with a minimum accepted error rate.

Activity 7: Data export - Data was exported in a text file format with a dictionary for further processing. This was the last step in the data processing system. The results were also exported in text files and their corresponding images of questionnaires to DVDs for backup and storage.

## 4. Platform Presentation

S2AEA is a Java platform dedicated to assessing strategic alignment using the concept of enterprise architecture. It contains two parts. The first part concerns the modelling of enterprise architecture and the second is dedicated to the strategic alignment evaluation.

The platform presented in this paper is the second version of S2AEA. The first version was presented in [4].

4.1 S2AEA v1

S2AEA (v1) is a Web oriented platform that provides interfaces describing enterprise architecture as a first step (fig3).

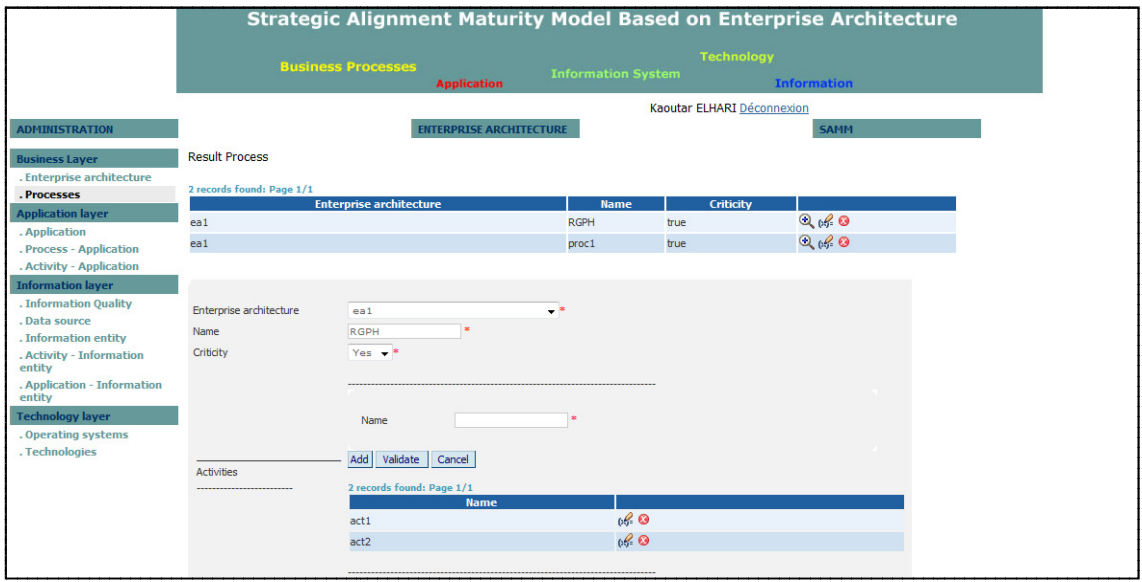

Fig3: Description of enterprise architecture using S2AEA v1

Strategic alignment maturity is calculated based on this description. Maturity tables are generated by corresponding to each layer's link, a level of maturity.

The approach here is interested globally in an alignment overview between layers. The figure 4 is an illustration of this approach.

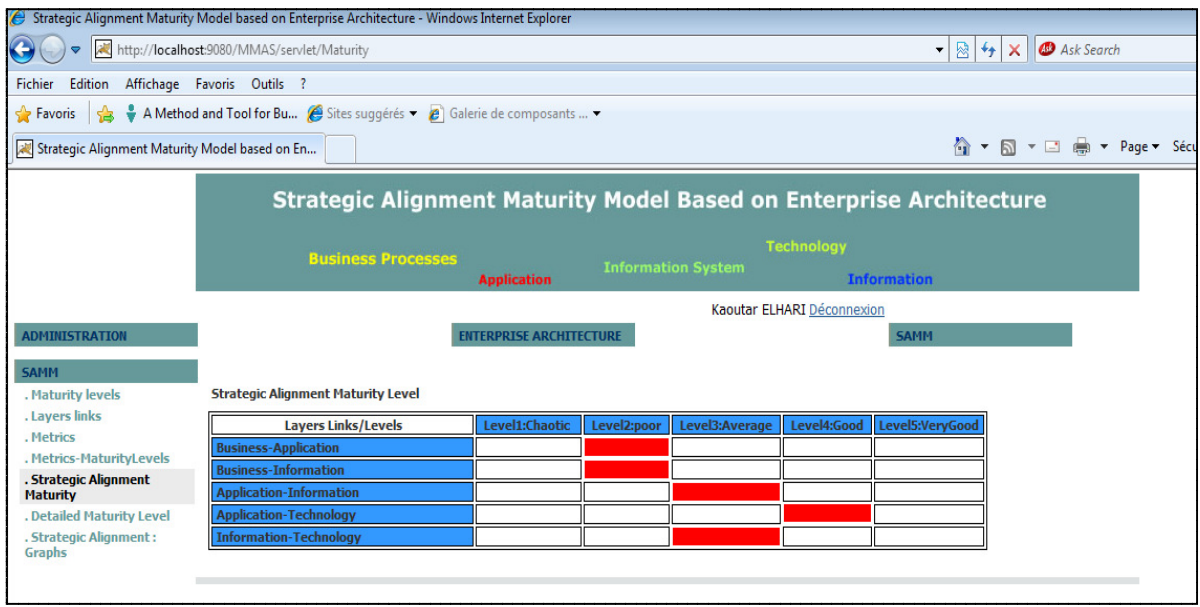

Fig4: Strategic alignment Maturity level using S2AEA v1

### 4.2 S2AEA v2

The version which is proposed in this paper offers the opportunity to shape the enterprise architecture graphically offering better ergonomics. The graphics incorporate the metamodel elements presented in fig3. The figure 5 illustrates how S2AEA (v2) models some activities of the process cited in section 3.

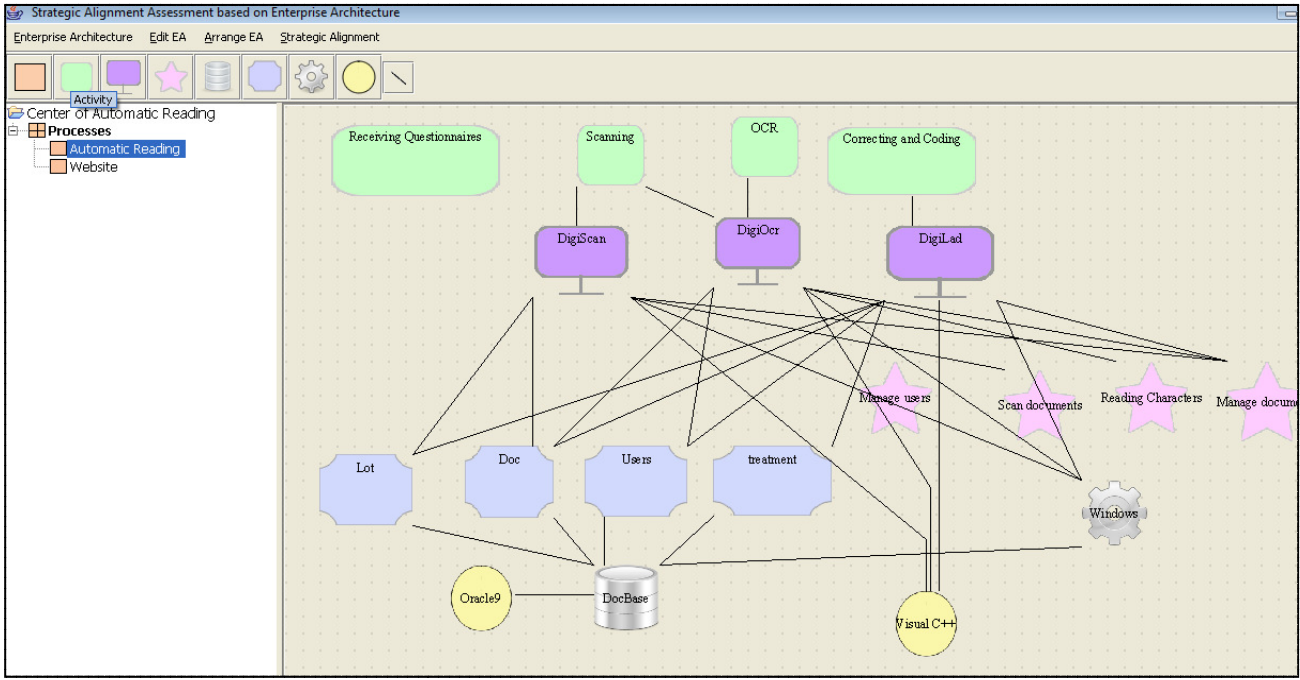

Fig5: Description of enterprise architecture using S2AEA (v2)

The table 1 contains elements constituting the figure5.

Table 1: S2AEA symbols

| Symbol | Name |
| --- | --- |
|  | Process |
|  | Activity |
|  | Application |
|  | Functionality |
|  | Data source |
|  | Information entity |
|  | Operating system |
|  | Technology |

The first version intends simply to calculate metrics and to infer the maturity level of each layer's link. It allows companies to locate their strategic alignment. S2AEA (v2) looks the alignment in more detail. It specifies information systems elements that affect the strategic alignment. This idea is based on 21 metrics collected in [4]. The v1 metrics targeted the whole layer while v2 metrics study case by case.

To illustrate an example of the use of S2AEA platform, we apply some metrics (M1 and M2) to the information system described in the figure5.

- M1: Number of activities not automated [4]

Indeed, each activity must be supported by an application in order to enhance alignment.

- M2: Number of applications supporting the same business process activity. [16], [18]

In fact, if a business process activity is supported by different applications; many problems can emerge:

- inserting the same data multiple times in different applications [21];
- Logging in multiple times, once for each application they need to access [21];
- etc

The figure 6 shows an example of two activities belonging to the process of automatic reading. It illustrates the role of the metrics M1 and M2.

After calculating metrics, the platform specifies the architecture elements that must be changed to reach a higher alignment level (activities red colored in figure 6).

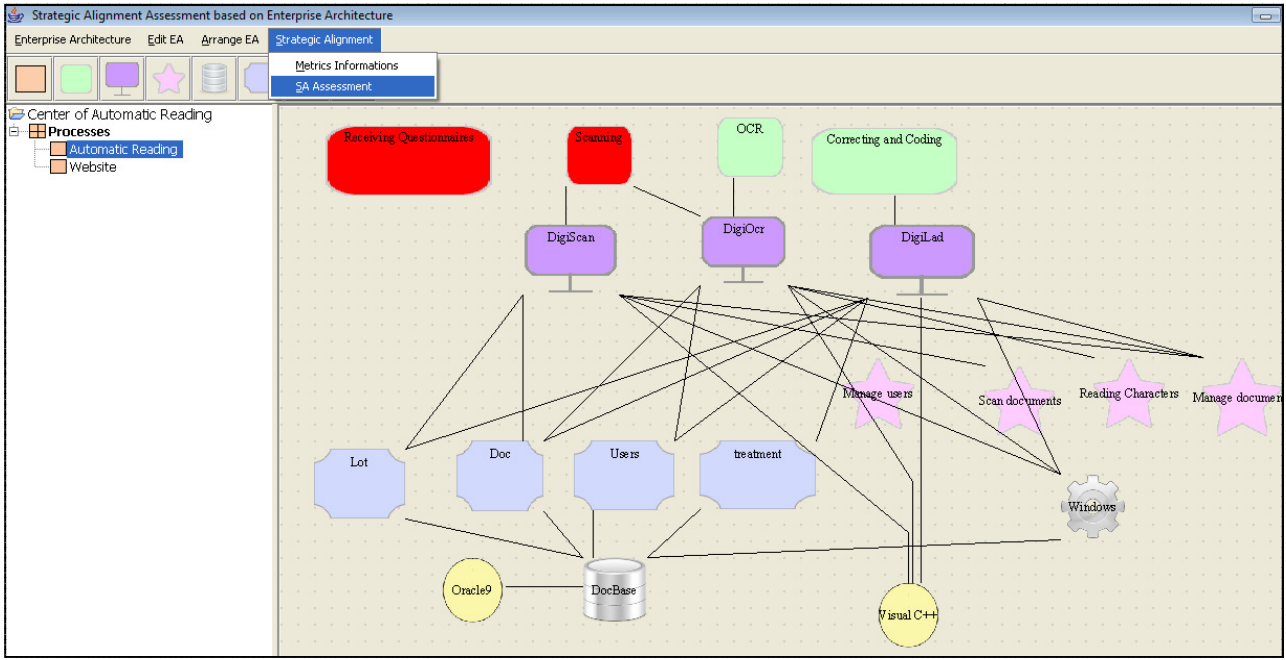

Fig 6: Strategic alignment assessment using S2AEA v2

The activity "Receiving questionnaires" harms the alignment in the sense that it is not automated (metric: M1). Architects should take it into account because it can be a real deficiency to deal with in order to reach alignment. Indeed, non automated activities require more human resources and more time. Figure 7 shows the message given by the platform concerning the activity "Receiving questionnaires".

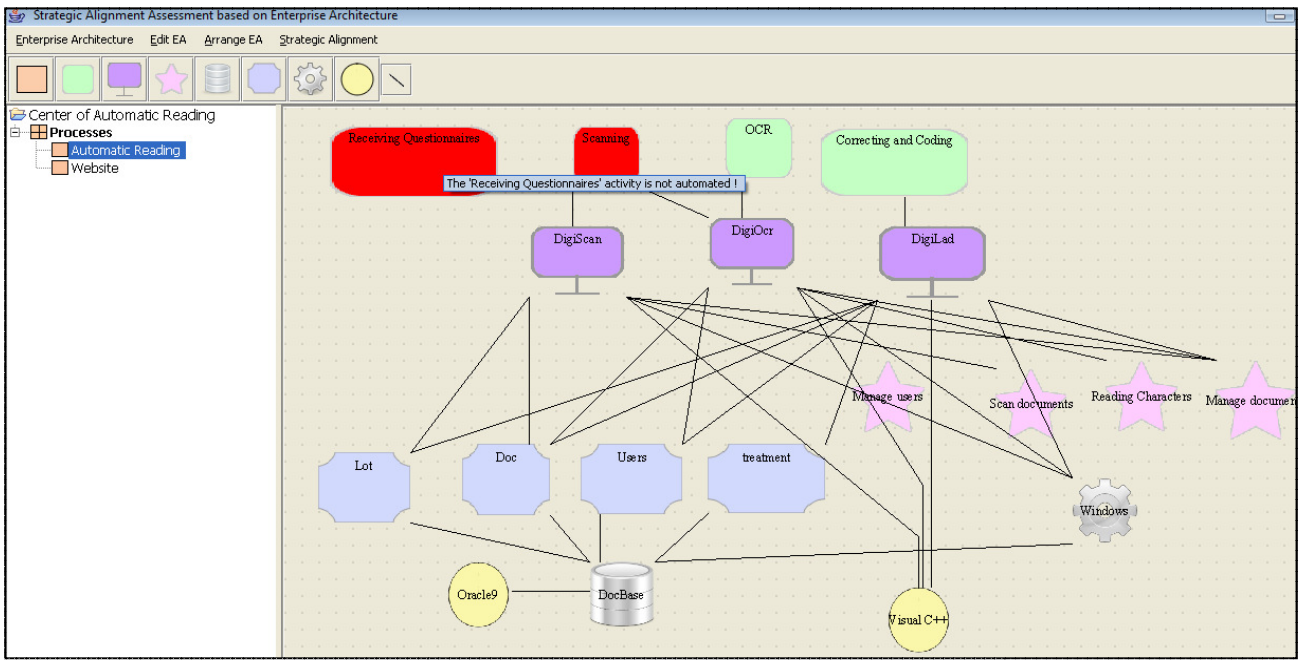

Fig7: Example of misaligned activity: activity not automated

On the other hand, figure 8 shows the problem raised by S2AEA concerning the "Scanning" activity. It takes into account the metric M2. "Scanning" activity harms the alignment because it is supported by three different applications (DigiScan, DigiOcr and DigiLad). Indeed, an activity must be supported by a minimum number of applications: This can facilitate modification when the business process activity changes [19] and can reduce the need for distributed transactions across applications [20] [21].

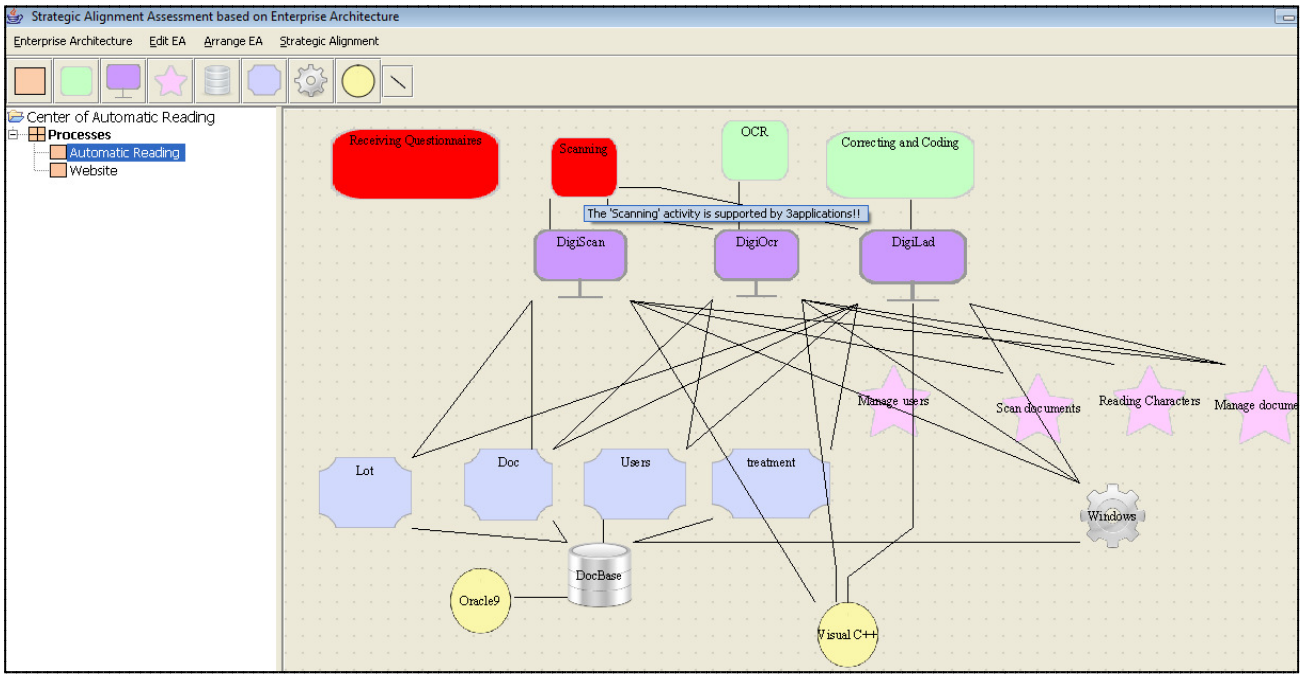

Fig8: Example of misaligned activity: activity supported by many applications

## 4. Conclusion

The article presents a platform S2AEA for assessing companies' strategic alignment.
The platform approach consists of using enterprise architecture concepts and its capacity to structure information system into layers. It is based on a set of metrics selected, studied and interpreted.
The platform proposed in this paper:
(a) graphically models enterprise architecture;
(b) calculates the corresponding metrics values;
(c) shows the information system elements harming strategic alignment;
(d) suggests the effort to do to reach a better strategic alignment level.
The very next steps in this research would be to improve S2AEA by adding more assessment metrics and by developing other platform functionalities.

**K. Elhari:** PhD candidate at the National High School for Computer Science and Systems Analysis (ENSIAS). She held an Extended Higher Studies Diploma from Mohammadia School of Engineers (EMI) on 2006 by working on the use of multi-agent systems in the development of Amine platform. She is a software Engineer graduated on 2003 from National Institute of Statistics and Applied Economics (INSEA) and she is working in the High Commission for Planning in the kingdom of Morocco.

**B. Bounabat:** PhD in Computer Sciences. Professor in ENSIAS, (National Higher School for Computer Science and System analysis), Rabat, Morocco. Responsible of "Computer Engineering" Formation and Research Unit in ENSIAS, Regional Editor of Journal of Computing and Applications, International Expert in ICT Strategies and E-Government to several international organizations, Member of the board of Internet Society - Moroccan Chapter.